# Return spillovers around the globe: A network approach

Štefan Lyócsa*

University of Economics in Bratislava, Institute of Economics and Management, Košice, Slovakia

Tomáš Výrost

University of Economics in Bratislava, Institute of Economics and Management, Košice, Slovakia

Eduard Baumöhl

University of Economics in Bratislava, Institute of Economics and Management, Košice, Slovakia

**Abstract**

Using a rolling windows analysis of filtered and aligned stock index returns from 40 countries during the period 2006-2014, we construct Granger causality networks and investigate the ensuing structure of the relationships by studying network properties and fitting spatial probit models. We provide evidence that stock market volatility and market size increases, while foreign exchange volatility decreases the probability of return spillover from a given market. We also show that market development and returns on the foreign exchange market and stock market also matter, but they exhibit significant time-varying behaviour with alternating effects. These results suggest that higher market integration periods are alternated with periods where investors appear to be chasing returns. Despite the significance of market characteristics and market conditions, what in reality matters for information propagation is the temporal distance between closing hours, i.e. the temporal proximity effect. This implies that choosing markets which trade in similar hours bears additional costs to investors, as the probability of return spillovers increases. The same effect was observed with regard to the temporal distance to the US market. Finally, we confirm the existence of the preferential attachment effect, i.e. the probability of a given market to propagate return spillovers to a new market depends endogenously and positively on the existing number of return spillovers from that market.

# Introduction

Understanding the transmission mechanism of stock market returns "spilling over" among international markets is crucial to quantify risk in financial decision making, both for investors and policy makers. In a current state of economic and financial liberalization it is widely believed, that equity market returns tend to move together and many researchers are focusing to quantify the extent of stock market co-movements around the world. The underlying idea is that high correlations among equity returns increase the overall risk of the investors' portfolio, thus if stock markets are highly-integrated, they provide only limited opportunities to diversify idiosyncratic risks effectively.

Since early works of Grubel (1968) and Solnik (1974) many markets implemented financial liberalization policies during the last few decades to become more integrated; but on the other hand, they also became more vulnerable to international risks and shock propagation. Benefits from international stock market integration has been challenged by many researchers using a wide spectrum of methodology, from simple Granger causality tests (Arshanapalli and Doukas, 1993), through correlations (Longin and Solnik, 2001), co-integration techniques (Mylonidis and Kollias, 2010), to various form of multivariate GARCH models (Cappiello et al., 2006), copula models (Aloui et al., 2011), to the most recent stream of methodology based on the variance decomposition from approximating models (Diebold and Yilmaz, 2014).

Although studies differ in so far they are using different methodologies, sample periods, study different markets or use other data sampling frequencies, the general consensus tends strongly towards strengthening of stock market linkages among most of world's equity markets. To mention few recent examples of observable return spillovers we could mention the Greek crisis or the meltdown of the Chinese stock market in 2015.

Our motivation for this paper can be best described using Figures 1a–1c, particularly 1b and 1c. The latter two figures show sub-networks, where an oriented edge between markets is created, if returns on one market influences (in Granger sense) returns on another market. In Figure 1b only effects of the US market are shown and it is clear, that the US market returns affect other returns in Asia, which open first in the next calendar trading day, but the effects last also for European markets, which open later the next day. However, Figure 1c is more interesting. Here only effects on the US market are shown. It is not that surprising that many European markets affect returns in the US as these are developed markets which close their trading sessions before trading session closes in the US. In this regard Asian stock markets do not appear to be influential at all, even though they are much larger and more developed (South Korea, Hong Kong, Japan) than the emerging stock markets in South Africa, Brazil and even Argentina (a frontier market), which still influence the US stock market. It appears that return spillovers are not determined only by economic fundamentals. An explanation for this finding is, that in a given calendar day, trading sessions on Asian markets end much earlier than those in Europe, South Africa (similar time zone as for European markets), or South America. Thus information implied in returns of the later markets might be informative regarding the development on the US market. Figure 1c shows that this is indeed the case.

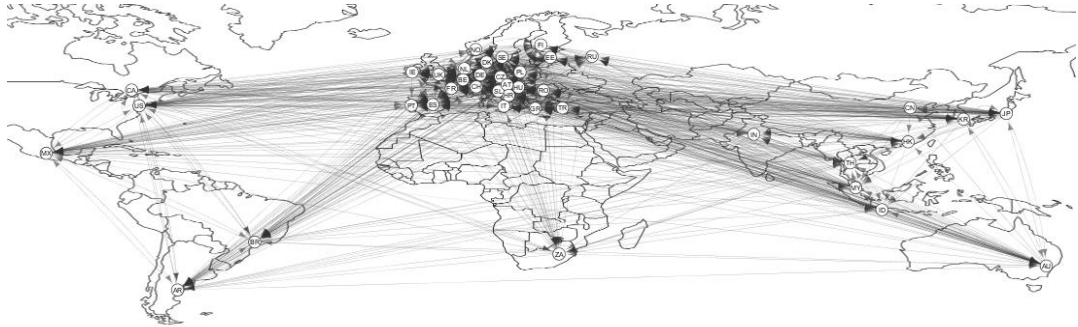

Figure 1a: Complex network of return spillovers during 2008

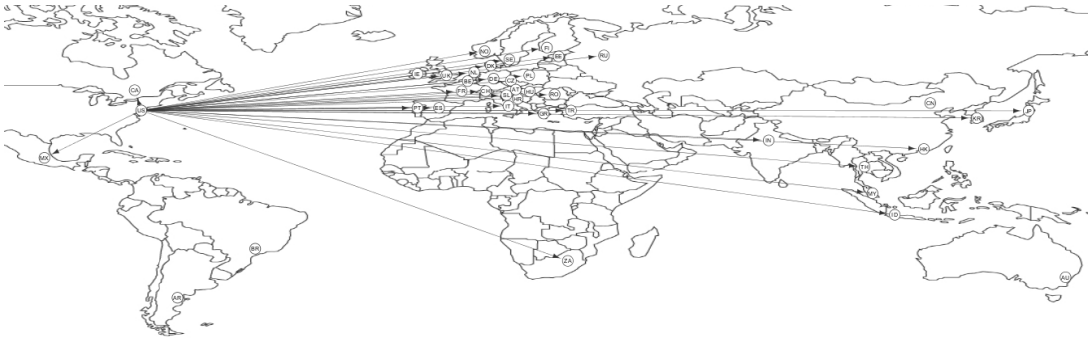

Figure 1b: Sub-network of return spillovers during 2008: US →
*Note: The network depicts only return spillovers from the US market to other markets around the world*

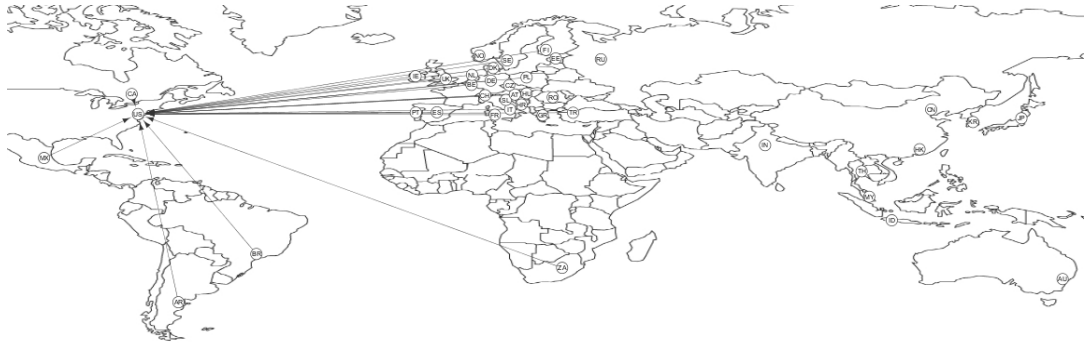

Figure 1c: Sub-network of return spillovers during 2008: → US
*Note: The network depicts only return spillovers to the US market from other markets around the world*

In this paper we present an approach, which allows us to quantify the effect of closing hours on the existence of return spillovers among equity markets, i.e. the temporal proximity effect (Výrost et al., 2015). Our first and main contribution is that we show that besides prevailing market conditions, the temporal proximity effect has a statistically and economically meaningful effect on return spillovers.

It is widely understood that the US market has large influence on the development of equity markets around the world. Our second contribution shows how the temporal proximity of other markets to the US market influences their ability to propagate returns to other markets.

Our third contribution is that we show that the probability of a given market to propagate return spillovers depends endogenously and positively on the number of return spillovers from that market. This effect is similar to the well-known preferential attachment effect described originally in network theory (e.g. Barabási and Albert, 1999).

Fourth, we contribute to the existing literature by showing that a large set of market related variables influence the probability of return spillovers in world equity markets. Finally, our descriptive and also econometric approach stems from the network perspective, rarely used in finance.

Using a sample of daily returns over the period from $2^{nd}$ January 2006 to $31^{st}$ December 2014 for 40 developed, emerging, and frontier markets, we test for Granger causalities among returns while controlling for the size of multiple Granger causality tests and taking care of return alignment with respect to non-synchronous trading effects. A possibly high number of return spillovers creates a complex network of relationships, which depicts world-wide market linkages. This is described via measures used in the network theory.

The remainder of the paper is organized as follows. In Section 1 we briefly introduce the reader to the networks and their use in finance. Section 2 describes the data, including the return alignment procedure used to deal with non-synchronous trading effects. Econometric and network methodology is described in Section 3. Section 4 describes the results and the last section concludes.

## 1 Networks and the stock markets

Since the influential paper of Barabási and Albert (1999), networks have penetrated many scientific domains, e.g. collaboration network of scientists or food web of marine organisms (Girvan and Newman, 2002), protein–protein interaction networks, metabolic networks, regulatory networks, RNA networks (Barabási et al., 2011), brain networks (Bullmore and Sporns, 2009), or other biological, social or technological networks (Dorogovtsev and Mendes, 2003). Networks have "infected" many fields, including finance and economics (Mantegna, 1999; Mantegna and Stanley, 1999), becoming an interdisciplinary approach (also a branch of science by its own) for problem solving.

Economic meaningfulness of correlation based networks[1] has been empirically demonstrated in many studies. For example, clustering of stocks from same industries was demonstrated in e.g. Onnela et al. (2003b), Tumminello et al. (2007), Tabak et al. (2010), Lyócsa et al. (2012). Clustering according to geographical proximity of markets have been found in Bonanno et al. (2000), Coelho et al. (2007), Gilmore et al. (2008), Eryiğit and Eryiğit (2009), Song et al. (2011). Changes in the structure of the relationships (i.e. topology of networks) during known crisis periods like Black Monday, currency crisis, dot-com bubble, recent financial crisis, US debt-ceiling crisis, or EU debt crisis, may be found in works of Onnela et al. (2003a), Song et al. (2011), Lyócsa et al. (2012), Trancoso (2014). Still, stock markets are rarely used in the mainstream finance and economics literature. Among few notable exceptions is the influential study of Billio et al. (2012) who constructed a graph of statistically significant Granger causalities among financial institutions. Further on, Diebold and Yilmaz (2014) employed their methodology to construct directional time-varying volatility networks of the US financial institutions.

The idea is to construct a network $G = (V, E)$, $V \subset \mathbb{N}$, where in our study vertices $V$ correspond to markets, and each edge $(i, j)$ from a set of edges $E$, where $E \subset V \times V$, corresponds to an interaction between two vertices $i$ and $j$. An interaction may be represented by a presence of Granger causality from vertex $i$ to vertex $j$. Such a network represents a structure of relationships between vertices. Using network specific indicators, one could answer empirically or theoretically motivated questions, e.g. does the changing structure of relationships precedes some economic events, when is the density of the network highest and why, how stable are relationships in networks, how are markets clustered?

---

[1] A short example could be as follows: A correlation matrix of equity returns is transformed using a suitable function to a distance matrix. Next, using a filtering method (the minimum spanning tree being the most widely used tool) a subset of most important correlations is selected. Based on the retained correlations an undirected network is created with weights corresponding to the distances (correlations). The resulting network structure then offers a simplified and often meaningful view at the complex relationships among returns.

The idea of creating Granger causality networks is certainly not new. We build upon the study of Výrost et al. (2015) who explored statistical properties of Granger causality networks created from 20 stock market indices. We differ in several ways. Most notably, we study not only topological properties and spatial factors as determinants of the resulting Granger causality networks, but are interested in various market conditions and market development characteristics. This allows us to explore the relative importance of these factors to each other. Further on we refined the methodology in several steps and our dataset includes not only updated data, but also additional 20 markets, which increased the number of potential linkages in a Granger causality network from 380 to 1560. Apart from the study of Výrost et al. (2015), lead-lag relationships for constructing networks were already exploited in the econophysics literature as early as in 2002 by Kullmann et al. (2002), and later used in Curme et al. (2014). Moreover, Granger causality networks were exploited also in the above mentioned study of Billio et al. (2012) and are a common tool to perform human brain mapping, e.g. Bullmore and Sporns (2009).

## 2 Data description and return alignment procedure

We study a sample of 40 market indices from five continents in a time period from $2^{nd}$ January 2006 to $31^{st}$ December 2014. According to the Dow Jones Classification System, 21 markets may be regarded as developed, 14 as emerging, and 5 as frontier markets. Data on annual market capitalization and market capitalization to GDP were obtained from World Development Indicators database of the World Bank[2]. Data on equity prices and exchange rates were obtained from the Thomson Reuters Datastream. The list of countries and stock market indices is available in Appendix A. Our sample of markets was chosen based on data availability of: (i) closing values, (ii) closing hours, and (iii) changes in closing hours. Our analysis of equity return spillovers is based on the local currency as we did not want to blur the extent of market co-movements with fluctuation on the foreign exchange market (Mink, 2015).

The Granger causality tests are based on a simple property that the past and present may cause the future but the future cannot cause the past (see Granger, 1969). It is therefore imperative to take into account closing hours of national stock markets. For each Granger causality test say from market $i$ to market $j$ ($i \neq j$) we have to align returns so that they correspond to the aforementioned principle[3]. We call this process return alignment rather than synchronization, as for almost all markets (except those which have same time-adjusted closing hours) returns cannot be synchronized at all (as they are non-overlapping).

---

[2] Basic data characteristics used in spatial probit models are presented in Appendix B.

Suppose we want to test for the presence of Granger non-causality between returns, $i \not\Rightarrow j$. Return alignment proceeds in the following three steps:

(i) List-wise deletion of stock prices is performed with respect to all missing (non-trading) days either on market $i$ or market $j$.

(ii) Next, for both markets, continuous returns $r_t = \ln(P_t/P_{t-1})$ are calculated, where $P_t$ denotes daily closing price at date $t$. The returns are calculated over all consecutive trading days; including returns over weekend, but returns over non-trading days during week are excluded.

(iii) The alignment of returns is performed in this step by considering closing hours at markets $i$ and $j$. In general, if we want to test for hypothesis $i \not\Rightarrow j$, we want to calculate correlation between returns on market $j$ and most recent but past returns on market $i$. For example, if market $i$ closes at 4:00 p.m. and market $j$ at 3:00 p.m. (time-zones adjusted), we use returns from market $i$ at $t-1$ to explain returns on market $j$ at $t$. Similarly, if market $j$ closes at 5:00 p.m, we now use returns from market $i$ at $t$ to explain returns on market $j$ at $t$. Without proper return alignment either we: (a) end up by testing $j \not\Rightarrow i$ instead of the intended $i \not\Rightarrow j$, or (b) we correlate returns on market $j$ at time $t$ using much older data on market $i$, which reduces our ability to find meaningful relationships.

We also have to take into account other sources of possible miss-alignment of returns:

a. We take into account changes in trading hours, specifically those related to the closing hours. It seems that as most studies which use daily data and perform some form of data synchronization report only the current closing hours. Possible historical changes in closing hours are not taken into account. This issue is important for Granger causality tests as some changes during analysed time periods lead also to different alignment of returns. For example, market $i$ might end its trading session before market $j$, but after the change in trading hours, market $i$ ends its session after market $j$ closes. One therefore needs to

---
[3] Symbol "$\not\Rightarrow$" denotes the Granger non-causality, i.e. should be read as "does not Granger-cause". Note, that for $j \not\Rightarrow i$ a different return alignment is necessary.

check for changes in closing hours and changes the return alignment process accordingly[4].

b. Some countries are not using daylight saving times (not to mention that some regions within a single country might use daylight saving time, while others might not). Some countries are determining daylight saving times on a year-to-year basis (e.g. Argentina). Moreover, the date of adjustment of time differs on a year-by-year basis and might not be the same for all countries. All these changes were taken into account as well.

c. It is not always straightforward to determine the exact time to which the last price belongs. Markets work with different types of closing auctions. For some markets, the price is not changing during the closing auction, only the quantity is determined. For some markets, the price might change during the closing auction, and/or the time to which the last price will belong is unknown in advance, as the time period for admitting orders is defined to be randomly determined on a day-to-day basis. In the latter case, we used closing time of the last possible trade, i.e. the hour at which the closing auction ends at latest. If closing auction was not based on the last known price during a regular trading session, we always tried using closing hours after the closing auctions.

## 3 Applied methodology

### 3.1 Granger causality test

We construct a network of return spillovers via Granger causality tests (for Granger causality see Granger 1969, 1980). At time $t$ information set of a time series $y_t$ is denoted as $I^y_t$. Similarly, for time series $x_t$ it is denoted as $I^x_t$ and $I_t = \{I^y_t, I^x_t\}$. We say that $x_t$ is Granger causing $y_t$ in mean, with respect to $I_t$ if:

$$E\left(y_t \middle| I^y_{t-1}\right) \neq E\left(y_t \middle| I_{t-1}\right) \tag{1}$$

In this paper we will utilize Granger causality test, initially proposed in Cheung and Ng (1996) as a test of Granger causality in variance. An adjustment of the test statistics for smaller samples is used as recommended by Hong (2001), and the test statistic will also take into account possible contemporaneous causality and will be calculated in rolling samples. The idea of the Cheung and Ng (1969) test is to test for the significance of the cross-lagged correlation coefficient of standardised conditional mean returns.

---
[4] Besides searching through home pages of stock markets and searching on the web, we double-checked our findings by contacting all stock exchanges in our sample. Exchanges which have not responded in the first survey have been contacted after one month again. Overall, we have identified 33 changes (extended or reduced trading hours) among all analysed markets during the examined period.

First, each series of returns $r_t$ is filtered via a suitable ARFIMA-GARCH model. The mean equation is defined as:

$$r_t = \alpha + z_t$$
$$\left(1 - \sum_{i=1}^{p} \phi_i L^i\right)(1-L)^d z_t = \left(1 + \sum_{j=1}^{q} \theta_j L^j\right)\varepsilon_t \qquad (2)$$
$$\varepsilon_t = \sigma_t \eta_t, \quad \eta_t \sim iid(0,1)$$

Where $\eta_t$ follows the Johson-SU distribution (Johnson, 1949a, b), with the probability density function:

$$f(x) = (2\pi)^{-1/2} J e^{-\frac{z^2}{2}} \qquad (3)$$

where $z = \varsigma^{-1}(sinh^{-1}(x) - \lambda)$ and $J = \varsigma^{-1}(x^2 + 1)^{-1/2}$. Here $\lambda$ and $\varsigma$ are parameters, which determine the skewness and kurtosis of the distribution. This choice of the distribution accounts for asymmetries and extreme tail market events (e.g. Choi and Nam, 2008).

Other nonlinearities can be captured by allowing variance $\sigma^2_t$ to be modelled by a GARCH process. Besides the standard GARCH model of Bollerslev (1986):

$$\sigma_t^2 = \omega + \sum_{k=1}^{r} \alpha_i \varepsilon_{t-k}^2 + \sum_{l=1}^{s} \beta_l \sigma_{t-l}^2 \qquad (4)$$

several other models were considered; AVGARCH (Taylor, 1986), NGARCH (Higgins and Bera, 1992), EGARCH (Nelson, 1991), GJR-GARCH (Glosten et al., 1993), APARCH (Ding et al., 1993), NAGARCH (Engle and Ng, 1993), TGARCH (Zakoian, 1994), FGARCH (Hentschel, 1995), CSGARCH (Lee and Engle, 1999). Both mean and variance equations were estimated within one single step – likelihood function. The entire analysis is conducted in R software using the rmgarch (Ghalanos, 2012a) and rugarch (Ghalanos, 2012b) packages.

For each series, a preferred specification was selected according to following steps. First, ARFIMA($p$,0,$q$)-GARCH($r$,$s$) models including all different variance equation specifications were estimated with all combinations of $p$, $q$, $r$, $s$ = 1, 2 and $d$ = 0. Second, only such specifications were retained, where the Peña and Rodríguez (2006) test with Monte Carlo critical values (see Lin and McLeod, 2006) suggested no autocorrelation and conditional heteroscedasticity in standardised residuals. These tests were performed for up to 20 lags in residuals, i.e. about one trading month. Third, we selected models with the lowest number of parameters (sum of $p$, $q$, $r$, $s$) as we preferred a more parsimonious representation. If more than one model remained (and this was often the case) the final model specification was selected according to the Bayesian information criterion (BIC, Schwartz, 1978). If however no suitable specification was identified, steps $1 - 3$ were performed as before, but now with $d \neq 0$. If still no model met our criteria, we selected the preferred specification from the set of all ARFIMA-GARCH models based on the BIC.

Suppose we test the null hypothesis of Granger non-causality from market $j$ to market $i$, $j \not\Rightarrow i$. Standardised conditional mean returns ($s_t = \varepsilon_t/\sigma_t$) from the preferred specifications are used to calculate the cross-lagged correlations:

$$\hat{\rho}(k) = \frac{\hat{C}_{ij}(k)}{\sqrt{\hat{C}_{ii}(0)\hat{C}_{jj}(0)}} \tag{5}$$

where

$$\hat{C}_{ij}(k) = \frac{1}{T}\sum_{t=k+1}^{T} s_{it} s_{jt-k}, k \geq 0 \tag{6}$$

It should be noted, that prior to the calculation of cross-lagged correlations, standardised conditional mean returns were aligned as specified in the next Section 2.[5]

The null hypothesis of Granger non-causality ($j \not\Rightarrow i$) is tested using the test statistic proposed by Hong (2001):

$$Q(M) = \frac{T\sum_{k=1}^{T-1} w^2(k/M)\hat{\rho}^2(k) - \sum_{k=1}^{T-1}(1-k/T)w^2(k/M)}{\sqrt{2\sum_{k=1}^{T-1}(1-k/T)(1-(k+1)/T)w^4(k/M)}} \tag{7}$$

Where we use the Bartlett weighting scheme:

$$w\left(z = \frac{j}{M}\right) = \begin{cases} 1-|z|, |z| < 1 \\ 0, |z| \geq 1 \end{cases} \tag{8}$$

In empirical simulation, Hong (2001) shows that the choice of $M$ and kernel function $w$ does not affect the size of the test[6], while power is affected only little. Under the null hypothesis, $Q(M)$ follows (asymptotically) the standardised normal distribution (it is a one-sided test). Note that (7) is calculated for a given (pre-determined) bandwidth $M$. We decided to use $M = 5$ as it corresponds to one trading week. A choice of $M = 3$ was also considered but led to almost identical results.

---

[5] Note also, that $k$ may sometimes (besides cases described by Eq. 9) be equal to 0 and still be valid for testing the hypothesis $j \not\Rightarrow i$. The minimum $k$ depends on the alignment of the standardised conditional mean returns (see Section 2).

[6] At least when a non-uniform weighting scheme is used, e.g. Bartlett or Quadratic Spectral.

For several markets (mostly in Europe) the time-zones adjusted closing hours are same. In these cases, we follow Lu et al. (2014) and allow for instantaneous return spillover from market $j$ to market $i$, by allowing $k = 0$ in calculating cross-lagged correlations, i.e.:

$$Q_u(M) = \frac{T\sum_{k=0}^{T-2} w^2(k/M)\hat{\rho}^2(k) - \sum_{k=1}^{T-1}(1-k/T)w^2(k/M)}{\sqrt{2\sum_{k=1}^{T-1}(1-k/T)(1-(k+1)/T)w^4(k/M)}} \qquad (9)$$

We tested for the presence of Granger causality in returns for all possible pairs within our samples. That led to 1560 statistical tests and a possibly high error rate. We decided to err rather on the side of safety and employed the rather conservative Bonferroni adjustment by using the significance level $0.01/(N(N-1))$, where $N$ is the number of stock markets. To achieve time variation we have applied the above procedure for rolling sub-samples of 12 months. The choice of 12 months is arbitrary and reflects our desire to be able to test for possible effects of some economic variables (which are available with annual data frequency) in the spatial probit models described below. The drift parameter is equal to 1. Over our sample from January 2006 to December 2014 we obtained 97 sub-samples. This approach is similar to that presented in Lu et al. (2014).

### 3.2 Stock market network modelling

Instead of calculating Granger causality tests for a small set of markets, we perform the analysis on a set of 40 markets. This creates a rather complex system of relationships. We use a graph as a mathematical construct to extract meaningful information, such as, e.g. which markets are most connected to other markets? How stable are these relationships over time?

Formally, define a directed graph $G_t = (V, E_t)$ at time $t$, with vertex set $V \subset \mathbb{N}$ corresponding to individual indices. The set of edges $E_t \subset V \times V$ contains all edges $(i, j)$ for indices $i, j \in V$ for which $i => j$, i.e. a directed edge from market $i$ to market $j$ is constructed if at a given Bonferroni adjusted significance level, return residuals on market $i$ Granger-cause return residuals on market $j$.

Probably the simplest measure of assessing the importance of a market within a network is to calculate its degree. The in-degree deg$^-$($i$) is defined as the cardinality:

$$\deg^-(i) = |\{(j, i) \in E_t; j \in V\}| \qquad (10)$$

Similarly, the out-degree is defined as:

$$\deg^+(i) = |\{(i, j) \in E_t; j \in V\}| \qquad (11)$$

The concept of a vertex degree as a measure of structural importance can be seen also from the fact, that it is equivalent to the so called "degree centrality". The central vertex is defined as the vertex with the highest vertex degree. Similar measures were used also in Billio et al. (2012) who used the degree of Granger causality as a ratio of the sum of edges to all possible edges and number of connections (standardised in- and out-degrees).

Degree of a market is a local measure, as it takes into account only its immediate connections. Billio et al. (2012) also used a global measure of centrality, namely the closeness, but it is not suitable for graphs which are not strongly connected (segmented markets without any relationships to other markets) or graphs with unreachable nodes (markets which are Granger causing other markets, but are not caused by other markets). Harmonic centrality is a relatively new measure which avoids the aforementioned pitfalls. Following Boldi and Vigna (2014), harmonic centrality of market $i$ can be defined as:

$$H(i) = \sum_{d(i,j)<\infty, i \neq j} \frac{1}{d(i,j)} \qquad (12)$$

where $d(i,j)$ is the shortest path from vertex $i$ to vertex $j$. If no such path exists, $d(i,j) = \infty$, we set $1/d(i,j) = 0$. The higher the market's harmonic centrality, the more central is the market within the given network, or to put it differently, the more important is the market for the flow of information.

The concept of vertex's centrality can be further adapted to the whole network. We will refer to such measures as centralization. According to Freeman (1979), there are two completely different approaches to what should be understood by network's centralization. The first approach is based on the notion of network's compactness, which leads to centralization measures which attempt to measure how close are vertices to each other. Two such measures of centralization are used in this study. The standardised average out/in degree:

$$\frac{1}{(|V|-1)|V|} \sum_{i \in V} \deg^+(i) \qquad (13)$$

The average harmonic centrality:

$$\frac{1}{|V|}\sum_{i \in V} H(i) \tag{14}$$

The second approach is based on the intuition of a relative difference between the most central vertex to all other vertices in the network, i.e. the more centralized network is more dominated (in terms of centrality) by one or a small group of vertices. For this purpose we have used the out-degree and in-degree centralization:

$$\frac{\sum_{i \in V}\left(\max_{j} \deg^{+}(j) - \deg^{+}(i)\right)}{(n-2)(n-1)} \tag{15}$$

$$\frac{\sum_{i \in V}\left(\max_{j} \deg^{-}(j) - \deg^{-}(i)\right)}{(n-2)(n-1)} \tag{16}$$

Finally, the stability of the network is considered using survival ratios as in Onnela et al. (2003b), which are simply the ratio of surviving edges. Refer to $E_t$ as a set of edges of the Granger causality network at time $t$. One-step survival ratio at time $t$ is defined as:

$$SR(s=1,t) = \frac{|E_t \cap E_{t-s}|}{|E_{t-s}|} \tag{17}$$

Multi-step survival ratio at time $t$ is then:

$$SR(s,t) = \frac{|E_t \cap E_{t-1} ... \cap E_{t-s}|}{|E_{t-s}|} \tag{18}$$

where, $s$ is the number of steps.

### 3.3 Spatial probit

As we consider each edge to signify the presence or absence of a relationship, it is interesting to analyse the characteristics that are related to the creation of edges. For example, is it more likely that returns on indices on larger markets tend to Granger-cause returns on other market indices? What other factors help explain the creation of edges?

The modelling of the existence/non-existence of an edge in a network naturally leads to a logit/probit type of model, with a binary dependent variable. We replicate the spatial probit approach used in Výrost et al. (2015). As we consider all possible edges within a network at the same time, some issues arise. For example, it is reasonable to assume some clustering of edges might be present. The probability of creating an edge between any two markets might therefore depend on the nature of vertices and thus the number of their existing linkages. This dependence raises some endogeneity issues with the modelling of the edge creation – clearly, the individual edges cannot be treated as independent of each other. To remedy this problem, we estimate spatial probit models proposed by McMillen (1992) and LeSage (2000), which take into account the interdependence between edges (for an overview of spatial models see LeSage, 2010).

To construct the model, we first define the dependent and independent variables. In our setting the variable of interest corresponds to the existence of links between the given nodes. We set $e_{ijt} = 1$ if $(i, j) \in E_t$, otherwise $e_{ijt} = 0$. We call **E** the matrix of all edge indicators $e_{ijt}$. To obtain our dependent variable (designated as **y**), we first vectorise the matrix of edge indicators (by calculating vec(**E**)), and then exclude the elements corresponding to the diagonal of **E**, as we are not interested in modelling loops – these have no economic meaning in our Granger analysis. We thus obtain a vector **y** of length $N(N-1)$.

Next, we define the matrix of spatial weights to indicate neighbouring observations, allowing for the modelling of spatial dependence. In our case, we have to define the spatial weight matrix **W** for all potential edges in **y**, thus **W** is a matrix of order $N(N-1) \times N(N-1)$. In general, for any two distinct possible edges $(i, j) \in V \times V$ and $(k, l) \in V \times V$ we set the corresponding element of **W** to 1 if the possible edges share the outgoing or incoming vertex (either $i = k$ or $j = l$)[7], 0 otherwise.

The spatial lag model (SAR) takes the form (LeSage, 2000, 2010):

$$\mathbf{y}^* = \rho \mathbf{W} \mathbf{y}^* + \mathbf{X}\boldsymbol{\beta} + \boldsymbol{\varepsilon}, \quad \boldsymbol{\varepsilon} \sim \mathbf{N}\left(\mathbf{0}, \sigma_\varepsilon^2 \mathbf{I}_{N(N-1)}\right) \tag{19}$$

Here the $\mathbf{y}^*$ represents an unobserved latent variable, which is linked to our variable of edge indicators **y** by:

$$y_i = \begin{cases} 1, & y_i^* \geq 0 \\ 0, & y_i^* < 0 \end{cases} \tag{20}$$

for $i = 1, 2, \ldots, N(N-1)$

As can be seen from (19), the existence of an edge is modelled by the existence of other neighbouring edges, as defined by the nonzero elements of matrix **W**, as well as exogenous variables **X**. The model parameters include the vector **β**, as well as a scalar $\rho$, which is related to spatial autocorrelation.

---

[7] For the purposes of estimation, we have used the row standardised version of **W** where the sum of elements in each row is equal to 1.

# 4 Empirical results and discussion

## 4.1 Granger causality networks

Causality tests described in Section 3.1 resulted in 97 sub-samples (from December 2006 to December 2014) of Granger causality networks. Each sub-sample is used to study return spillovers over the past 12 months. The maximum number of return spillovers is 1560. In our empirical application the mean value was 823 (which corresponds to 52.8% or 0.528 in mean out/in-degree centrality, see Figure 2). Even the minimum of 532 return spillovers means, that visualizing the full Granger causality network is not very informative due to its complexity (see example in Figure 1a). Instead, in Figure 1b and Figure 1c we selected sub-networks from a sub-sample of return spillovers over the year 2008. More specifically, Figure 1b depicts only spillovers from the US market, while in Figure 1c spillovers to the US market are shown. An interesting observation is that while stock markets in Asia are not influencing the US market, markets in Europe, South Africa (similar time-zone to that of many European markets) and even frontier market of Argentina is Granger causing returns in the US. This observation represents our main motivation for this study as it clearly shows that return spillovers among markets are at least partially influenced by relative setting of trading hours of a given markets, i.e. the temporal proximity. We will explore this more formally in Section 4.3.

Figure 2 plots four measures of spillover connectedness which are assigned to the last month of a given sub-sample. We observe three different aspects of return spillover development over the sample period. The out-degree and in-degree centralization show, that in the past 12 months prior to December 2007 (out-degree) and January 2008 (in-degree), there were few markets which influenced many others (a peak in out-degree centralization), while at the same time, many markets were influenced in the same extent (a bottom in in-degree centralization). These properties seem to show that there are market conditions on world equity markets, when few markets influence all the others.

We were interested whether this can be related to the financial crisis and the US market. A closer examination revealed that at least in the Granger sense, US market appeared to be much more influential only since October 2008 (after the Lehman Brothers collapse) up until the end of June 2009. Out of 39 markets, during that period, the US market was influencing in average 33.44 markets (max 37 during 12 months ending in December 2008).

Mean out/in-degree centrality and mean harmonic weighted centrality (lower panel of Figure 2) reflect a very similar but perhaps surprising development of return spillovers. Both measures attempt to quantify the compactness of the return spillover network, i.e. the higher values are associated with more interconnected markets. However, we can observe, that since May 2012 there is a sharp drop in interconnectedness of markets around the world. Before that period, values of around 0.60 in mean out/in-degree centrality mean that among all possible spillovers (1560) around 60% were statistically significant (936); after that, these numbers dropped to as low as 34% (530). Unfortunately, our sample starts only in the end of 2006, and we are therefore unable to confirm, whether since 2007 (start of the crisis on equity markets) up until May 2012 we have not observed a period of higher stock market interlinkages.

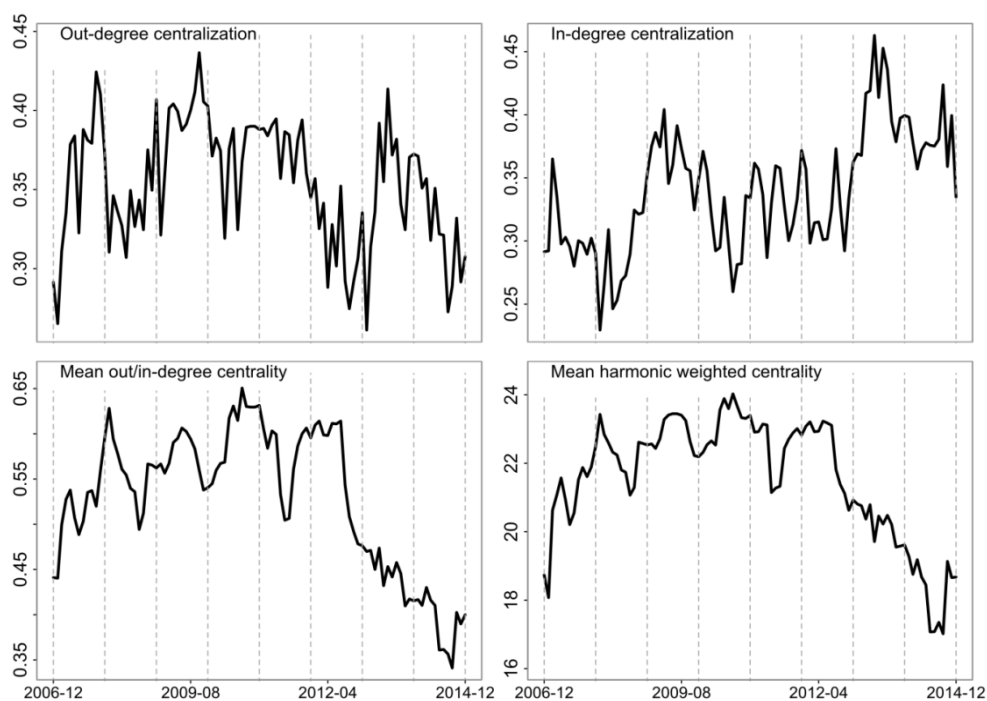

Figure 2: Time-varying spillovers: network centralization

## 4.2 Connectedness of markets: A network approach to return spillovers

Table 1 reports descriptive statistics of out- and in-degrees and harmonic centralities of individual markets. The former two are local measures of market's connectedness as it is simply the average number of direct spillovers from (out-degree) and to the selected market (in-degree), i.e. connections to neighbouring vertices in a network. The lowest out-degree is found to be for Portugal (7.454 return spillovers in average), Slovenia (8.175) and China (9.948). While the former two are smaller markets situated in Europe, thus they close trading around the same time as most of markets in Europe, the stock market in China closes much earlier. Normally, one could therefore expect that it will lead European markets, which in turn would lead to higher out-degree. However, it seems that the stock market in China is rather separated from other equity markets around the world. This reasoning is apparently suggestive when one considers other Asian markets in the region, which in turn have one of the highest average out-degrees: South Korea (31.629), Hong Kong (30.546), and Japan (27.464). Average number of out-degree for UK and US is much lower at 16.031 and 19.186, respectively, but one should bear in mind, that our sample of markets is over-represented with European markets, which open their trading sessions much closer to closing times of Asian markets. This is therefore another example of possible temporal proximity effect[8].

With average of 4.454, the lowest in-degree was measured by far for China, which again strengthens our belief, that during our sample period, the stock market in China appears to be much more segmented compared to other equity markets in the world. Highest average in-degree was found for the UK market (31.340), but several other markets are also highly influenced by other markets: Canada (30.629), Portugal (30.577), Netherlands (30.433), and France (30.031). Interestingly, also markets in South America (Argentina and Brazil) were subject to return spillovers, although one would perhaps expect that, at least for Argentina, the market would act more segmented.

A further observation is, that the average number of out-degrees does not seem to be strongly positively correlated with the level of the development of the country, as several frontier/emerging markets have higher average out-degree than markets in developed countries (Romania 21.082 vs. Netherlands 16.247). Similarly, markets in developed countries seem to have higher number of in-degrees. This also suggests, that assumption that only economic fundamentals or behavioural factors influence return spillovers is false.

---

[8] Another non-contradicting explanation might be, that returns on smaller markets (in terms of market capitalization) with large share of global companies (e.g. companies of large conglomerates in South Korea like Samsung, LG and Hyundai) are more related to global economic development and local factors are not so important drivers of the returns of the local market index.

Table 1 Connectedness of markets: vertex centrality

|  | Out-degree | | | | In-degree | | | | Harmonic centrality | | | |
|---|---|---|---|---|---|---|---|---|---|---|---|---|
|  | Mean | SD | Max | Trend | Mean | SD | Max | Trend | Mean | SD | Max | Trend |
| *Frontier markets* | | | | | | | | | | | | |
| AR | 12.052 | 4.957 | 28 | -0.114 c | 28.340 | 6.305 | 38 | -0.131 a | 18.410 | 2.281 | 24.146 | -0.055 c |
| HR | 17.278 | 10.606 | 30 | -0.081 | 10.247 | 6.789 | 32 | -0.128 a | 18.455 | 7.694 | 25.995 | -0.058 |
| EE | 15.814 | 8.616 | 29 | -0.008 | 10.639 | 6.454 | 34 | -0.124 | 19.259 | 3.995 | 24.888 | -0.001 |
| RO | 21.082 | 8.833 | 32 | 0.083 | 11.856 | 5.331 | 30 | -0.053 | 20.497 | 6.825 | 26.356 | 0.085 |
| SL | 8.175 | 9.305 | 28 | -0.097 | 9.423 | 9.125 | 29 | -0.047 | 14.104 | 7.606 | 24.879 | -0.106 |
| *Emerging markets* | | | | | | | | | | | | |
| BR | 16.577 | 6.439 | 36 | -0.032 | 29.680 | 4.448 | 36 | -0.062 | 20.307 | 2.831 | 28.054 | -0.022 |
| CZ | 26.990 | 3.050 | 34 | -0.066 a | 16.557 | 3.916 | 27 | -0.027 | 24.179 | 1.559 | 27.050 | -0.034 a |
| HU | 22.897 | 4.462 | 31 | -0.082 | 14.247 | 3.669 | 21 | -0.089 c | 22.520 | 2.424 | 26.680 | -0.042 |
| CN | 9.948 | 7.650 | 28 | 0.026 | 4.454 | 3.577 | 19 | 0.001 | 17.279 | 4.172 | 24.635 | 0.018 |
| IN | 27.351 | 4.253 | 36 | -0.011 | 11.093 | 3.868 | 25 | -0.089 c | 24.149 | 1.963 | 27.615 | -0.010 |
| ID | 25.959 | 7.842 | 34 | -0.147 | 13.546 | 6.922 | 30 | -0.034 | 23.432 | 3.343 | 27.424 | -0.064 |
| MY | 24.412 | 9.336 | 37 | -0.225 c | 18.093 | 8.263 | 32 | -0.099 | 22.749 | 3.690 | 27.661 | -0.086 c |
| MX | 14.804 | 6.743 | 30 | -0.181 c | 29.526 | 3.467 | 36 | -0.037 | 19.542 | 2.848 | 25.781 | -0.076 c |
| PL | 24.381 | 3.701 | 33 | -0.075 b | 16.711 | 3.626 | 25 | 0.040 | 23.149 | 1.909 | 26.892 | -0.039 c |
| RU | 25.897 | 4.045 | 34 | -0.070 a | 14.093 | 3.929 | 23 | -0.055 | 23.661 | 1.812 | 26.467 | -0.029 a |
| ZA | 24.134 | 3.561 | 32 | -0.013 | 17.660 | 3.379 | 24 | -0.058 c | 22.927 | 1.837 | 26.520 | -0.013 |
| KR | 31.629 | 6.371 | 39 | -0.133 | 22.351 | 7.429 | 33 | -0.124 b | 25.703 | 2.578 | 28.978 | -0.051 |
| TH | 24.340 | 7.047 | 33 | -0.045 | 12.990 | 7.050 | 28 | -0.034 | 22.770 | 3.110 | 26.924 | -0.028 |
| TR | 26.464 | 4.535 | 36 | -0.127 c | 10.825 | 4.704 | 18 | -0.104 c | 23.815 | 2.189 | 28.001 | -0.064 c |
| *Developed markets* | | | | | | | | | | | | |
| AU | 29.557 | 6.699 | 39 | -0.069 | 23.897 | 5.380 | 30 | -0.057 | 24.999 | 2.715 | 28.971 | -0.030 |
| AT | 16.340 | 2.203 | 23 | -0.008 | 25.948 | 2.413 | 32 | -0.025 | 20.251 | 1.260 | 22.930 | -0.014 |
| BE | 15.959 | 2.250 | 23 | -0.021 | 29.887 | 3.182 | 35 | -0.060 c | 20.207 | 1.043 | 23.155 | -0.016 c |
| CA | 13.784 | 4.248 | 25 | -0.058 | 30.629 | 3.404 | 37 | -0.008 | 19.271 | 1.746 | 22.736 | -0.023 |
| DK | 23.546 | 2.890 | 29 | -0.040 | 17.845 | 2.506 | 24 | -0.046 b | 22.960 | 1.538 | 25.623 | -0.025 |
| FI | 22.918 | 2.853 | 30 | 0.010 | 23.670 | 2.120 | 27 | -0.035 | 22.926 | 1.334 | 25.742 | -0.004 |
| FR | 16.340 | 2.354 | 22 | -0.025 | 30.031 | 2.899 | 35 | -0.040 | 20.362 | 1.160 | 22.742 | -0.018 b |
| DE | 16.041 | 2.101 | 21 | -0.030 a | 29.784 | 2.232 | 33 | -0.016 | 20.256 | 1.142 | 22.297 | -0.022 c |
| GR | 23.351 | 6.487 | 36 | -0.177 a | 9.876 | 5.272 | 18 | -0.157 c | 22.724 | 2.936 | 28.190 | -0.081 c |
| IE | 22.196 | 3.226 | 28 | -0.037 | 22.485 | 2.582 | 28 | -0.028 | 22.494 | 1.640 | 25.056 | -0.024 |
| IT | 22.742 | 2.176 | 27 | -0.021 | 22.041 | 3.485 | 28 | -0.050 a | 22.819 | 1.227 | 25.350 | -0.017 |
| JP | 27.464 | 5.605 | 35 | -0.126 c | 22.485 | 8.800 | 34 | -0.204 | 24.158 | 2.214 | 27.650 | -0.050 c |
| NL | 16.247 | 2.806 | 22 | -0.039 | 30.433 | 2.056 | 34 | -0.011 | 20.278 | 1.506 | 22.784 | -0.026 b |
| HK | 30.546 | 4.330 | 36 | -0.061 | 19.763 | 7.143 | 30 | -0.047 | 25.377 | 1.939 | 27.823 | -0.027 |
| NO | 25.299 | 2.399 | 31 | 0.001 | 19.835 | 6.298 | 27 | -0.068 | 23.739 | 0.969 | 26.140 | -0.005 |
| PT | 7.454 | 2.273 | 13 | 0.009 | 30.577 | 2.922 | 38 | -0.032 | 16.518 | 1.364 | 19.634 | -0.003 |
| ES | 15.361 | 1.501 | 19 | -0.016 b | 28.495 | 2.190 | 33 | -0.033 b | 19.921 | 0.848 | 21.703 | -0.015 c |
| SE | 22.278 | 2.684 | 28 | -0.029 a | 22.845 | 1.954 | 26 | -0.016 | 22.625 | 1.387 | 24.984 | -0.022 c |
| CH | 21.093 | 2.420 | 25 | -0.017 | 22.979 | 2.165 | 28 | -0.014 | 22.074 | 1.342 | 24.090 | -0.019 a |
| UK | 16.031 | 2.800 | 23 | -0.032 | 31.340 | 1.978 | 35 | -0.014 | 20.173 | 1.541 | 22.962 | -0.024 b |
| US | 19.186 | 6.517 | 37 | -0.045 | 26.742 | 3.528 | 34 | -0.015 | 21.319 | 2.508 | 27.939 | -0.020 |
| MG | 20.598 | 6.059 |  | -0.056 c | 20.598 | 7.647 |  | -0.056 c | 21.559 | 2.522 |  | -0.029 c |

*Note: trend denotes the estimated trend coefficient of a simple linear time trend regression, where the dependent variable is out-degree (in-degree, or harmonic centrality) of a corresponding market.* **a**, **b**, **c** *denote statistical significance at the 10%, 5%, and 1% level, respectively. We have used the HAC Newey-West standard errors estimated with automatic bandwidth selection and quadratic spectral weighting scheme as in Newey and West (1994). MG corresponds to the pooled mean group estimator.*

Contrary to simple out/in-degrees, harmonic centrality is a global measure of market's connectedness as it also takes into account indirect spillovers in the full network. The idea is that even though market A might not Granger-cause returns on market C directly, it might do so through market B, if market B Granger-causes returns on market C and market A Granger-causes returns on market B. In our empirical application, the largest possible value for a harmonic centrality is 39 (star-like network). Therefore the harmonic centralities appear to be not only very similar among markets, but also rather large, which suggest that taking indirect spillovers into account, the connectedness of stock markets around the world is high. Such environment should be vulnerable to contagion.

Linear trend analysis confirms what was already visible from Figure 2, the mean out/in-degree centrality, that at least within our sample period, the number of return spillovers is slightly declining. Interestingly the role of the US market seems to be changing. The number of out/in-degrees has declined (see Table 1). It seems that the role of the US market has diminished, meaning that the US returns are less indicative about the development on other markets around the world. However, as pointed out by Výrost et al. (2015) it also might be, that market moving news are increasingly reported rather after-hours (perhaps to decrease the volatile response of markets), thus closing returns are not reflecting the after-hours news. If this is true, then the temporal proximity of closing hours relative to those of the US market should be indicative with respect to the occurrence of return spillovers. The closer the market is to the closing hours of the US market, the more likely it should be that that market will influence others, as this market will act like a hub of after-hours news on the US market. We will test this hypothesis in Section 4.3.3.

Based on average out/in-degree, we have also observed some tendency of markets with higher out-degree to have a rather lower in-degree (Pearson's correlation –0.29, left panel in Figure 3). If there would be no temporal proximity effects, it would suggest that there are markets which rather than being influenced tends to influence others, i.e. market-moving markets. However, if we assume that the temporal proximity effects are fixed[9], the time-variation of these out/in-degree correlations can be interpreted as a decline, means that some markets have increased their influence on other markets in the network. These correlations are plotted in the right panel of Figure 3 and clearly show that this is what happened during the recent financial crisis which originated in the US. The sharp drop actually corresponds to the sub-period ending in September 2008, i.e. the data cover the period from October 2007 until September 2008.

---

[9] This seems to be reasonable as the closing times change rarely.

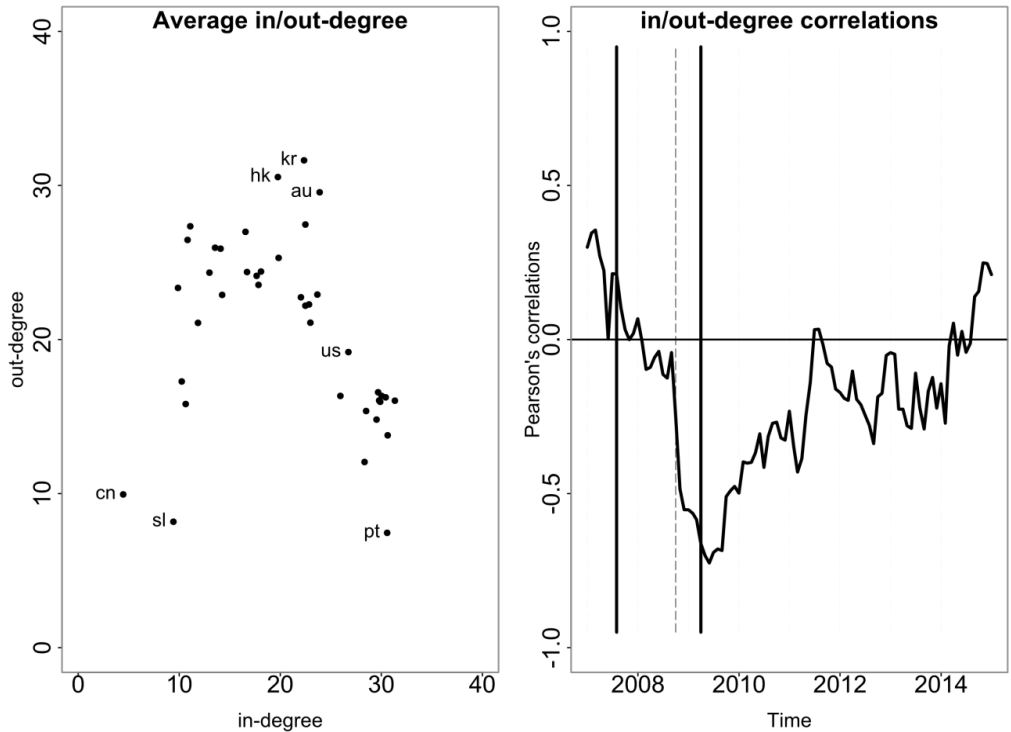

Figure 3: In-/out-degree relationship.
*Note: The left panel is a scatterplot of average in- and out-degrees. The right panel is a time series of in-/out-degree correlations calculated for each of the 97 sub-samples. The vertical solid lines are defining the crisis period (from July 2007 until March 2009), while the dashed vertical line corresponds to the sub-sample ending in September 2008.*

Finally, we are also able to answer a question about the stability of equity return spillovers by calculating survival ratios. If return spillovers show higher resiliency to changing market conditions, it has implication for the stability of all policies which rely on international market co-movements, e.g. monitoring policies regarding the fragility of financial markets or international investors. The left panel in Figure 4 shows that return spillovers are very stable, for example after one month (when one month of data are removed and a new month of data are added to the sub-sample), 95.6% of return spillovers have survived. This might be attributed to the fact, that we are using overlapping sub-samples, as our sub-samples of length 12 months are rolled one month ahead. However, even after 12 months, there are still 74.84% surviving return spillovers. In general, it therefore seems that the structure of return spillovers among equity markets is stable.

As these survival ratios are all averaged over time, the right panel in Figure 4 shows also the time variation of selected survival ratios. While no apparent changes are observed for single step survival ratio of return spillovers, the lower right panel of Figure 4 shows that at the end of our sample period, the ratio of surviving return spillovers was below 70%, but still over 50% spillovers are surviving even after 12 months. The next section explores the determinants of return spillovers and thus in some respect also the stability of return spillovers over time.

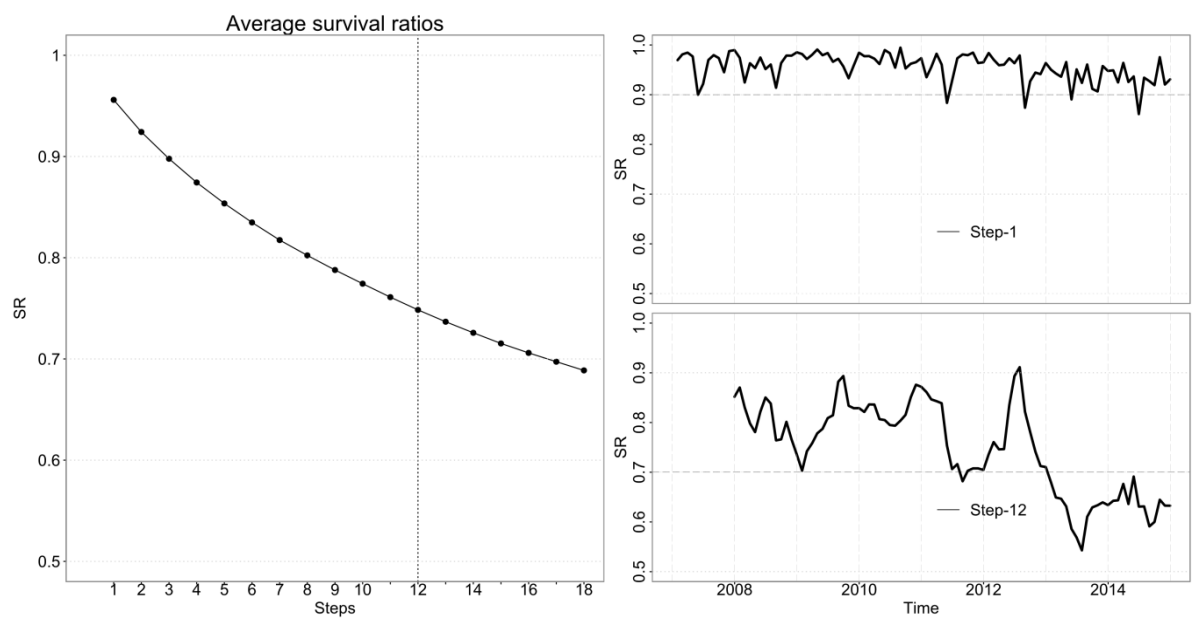

Figure 4: In-/out-degree relationship.
*Note: The left panel denotes the average ratio of surviving return spillovers after x number of months. The right panel denotes the time variation of a ratio of surviving return spillovers after one month (upper right figure) and after 12 months (lower right figure).*

## 4.3 Determinants of market's connectedness

To analyse and explain the formation of edges within the network structures, we have fitted spatial autoregressive probit models. The binary dependent variable denotes the presence/absence (1/0) of a directed edge, i.e. a statistically significant return spillover from one market to another. One has to note that the edges are oriented, as Granger causality is directional. Hence, we distinguish between in- and out-vertices, with the out-vertex Granger causing the in-vertex. The information flow is thus from the out-vertex to the in-vertex. The independent variables in model (19) are all related to the out-vertex. We included returns on the equity market index over the corresponding rolling window, FOREX returns (to USD), realized volatility on both equity and FOREX markets, market capitalization and market capitalization to GDP. Both the index and FOREX returns have been standardized over the full sample to have zero means and unit variances, in order to keep the returns (and the estimated coefficients) from various markets comparable. The standardization also affects the magnitude of the calculated realized volatilities. Separate dummy variables have been included to describe, whether the out-vertex market is a developed market, while the potentially in-vertex market is either an emerging market (one dummy) or a frontier market (second dummy). The descriptive statistics of these variables are presented in Appendix B.

We also included several spatial factors, related to the position of individual markets within our network and within time zones and trading hours. Specifically, we have used the time difference between markets (measured as described in section 2), time difference to the US market to assess its dominance within the analysed group of markets, as well as the spatial autocorrelation coefficient, which indicates the presence of spatial dependence. The time differences have been calculated in a way that ensures the non-negativity of all values, as the difference was always measured as the amount of time from market close of the out-vertex to the preceding market close of the in-vertex.

To capture the dynamics within the networks, the estimation was conducted on rolling windows spanning 12 months, with drift of 1 month. Values of all variables have been set as of the last day of the rolling window. The average values of the coefficients, frequency of occurrence of positive and negative values, as well as the number of significant coefficients for each explanatory variable are shown in Table 2.

Table 2 Average spatial probit coefficients

|  | Average coefficient | positive coefficient # | positive coefficient # signf. at 0.05 | negative coefficient # | negative coefficient # signf. at 0.05 |
|---|---|---|---|---|---|
| *Panel A: spatial factors* |  |  |  |  |  |
| Spatial coefficient | 0.7535 | 97 | 96 | 0 | 0 |
| Temporal distance | -0.0025 | 0 | 0 | 97 | 97 |
| Temporal distance to US | -0.0009 | 0 | 0 | 97 | 84 |
| *Panel B: market factors of out-vertex markets* |  |  |  |  |  |
| Return on equity market | 0.1511 | 52 | 31 | 45 | 30 |
| Volatility on equity market | 0.4340 | 78 | 47 | 19 | 3 |
| Return on FOREX | 0.4909 | 57 | 27 | 40 | 13 |
| Volatility on FOREX | -0.3105 | 7 | 1 | 90 | 57 |
| Market capitalization | 0.0949 | 97 | 96 | 0 | 0 |
| Market capitalization to GDP | 0.0017 | 55 | 22 | 42 | 17 |
| Developed to frontier market | 0.1687 | 65 | 37 | 32 | 13 |
| Developed to emerging market | 0.0679 | 65 | 13 | 32 | 3 |

*Note: We report the average coefficient calculated across all rolling-windows. Next we report the number of times the given coefficient was positive and a number indicating in how many instances positive coefficients were also significant at the 5% significance level. The same is reported for negative coefficients.*

### 4.3.1 Are market conditions influencing return spillovers?

From Table 2 it is clear that there are many periods, when returns and volatility on equity and FOREX markets matter. Regarding the returns, the results might appear mixed. With integrated markets, one would expect a positive sign on the return coefficient. However, our sample covers rather heterogeneous group of markets, which might lead to flights of investors from one market (selling) to another (buying) thus causing negative signs. It turns out, that at least within our sample, sometimes this was the case. Positive coefficients are in accordance with numerous studies, which found evidence either for periods of increased co-movement, due to contagion or impact of common factors. On the other hand, negative coefficients are in-line with the return chasing hypothesis, when investors tend to leave markets where returns are expected to be low and move to markets with higher expected returns potential (Bohn and Tesar, 1996). Results from our rolling-window analysis suggest, that for a given period one motive out-weights the other, thus for a given sample of periods both motifs might co-exists.

Effects of the returns on the foreign exchange also show visible time-varying behaviour. Significant and positive coefficients outnumber negative and significant coefficients two to one. The relationship between stock and foreign returns is not a one that is described easily, as current theoretical approaches allow for relations in both directions (e.g. Hau and Rey, 2006; Cenedese et al., 2015). In our case, the situation is more complicated, because our dependent variable models the Granger causalities between stock returns, not the returns themselves. For example, if one accepts the portfolio balance model of the relationship between stock market and foreign exchange rate, then decreases on the local stock market should lead to depreciation of the local currency and based on our results, most of the time this should increase the probability of a return spillover from local to a foreign market. However, for some periods, the decrease-depreciation situation leads to decrease in the probability of return spillovers. However, compared to equity returns, the effect of foreign exchange returns seems to be stronger (in Figure 5 axes are equally scaled). This might be the consequence of many emerging and frontier markets in our sample, where foreign exchange returns (over a 12 month period – the length of our sub-samples) might represent a significant part of the overall return of international equity investors.

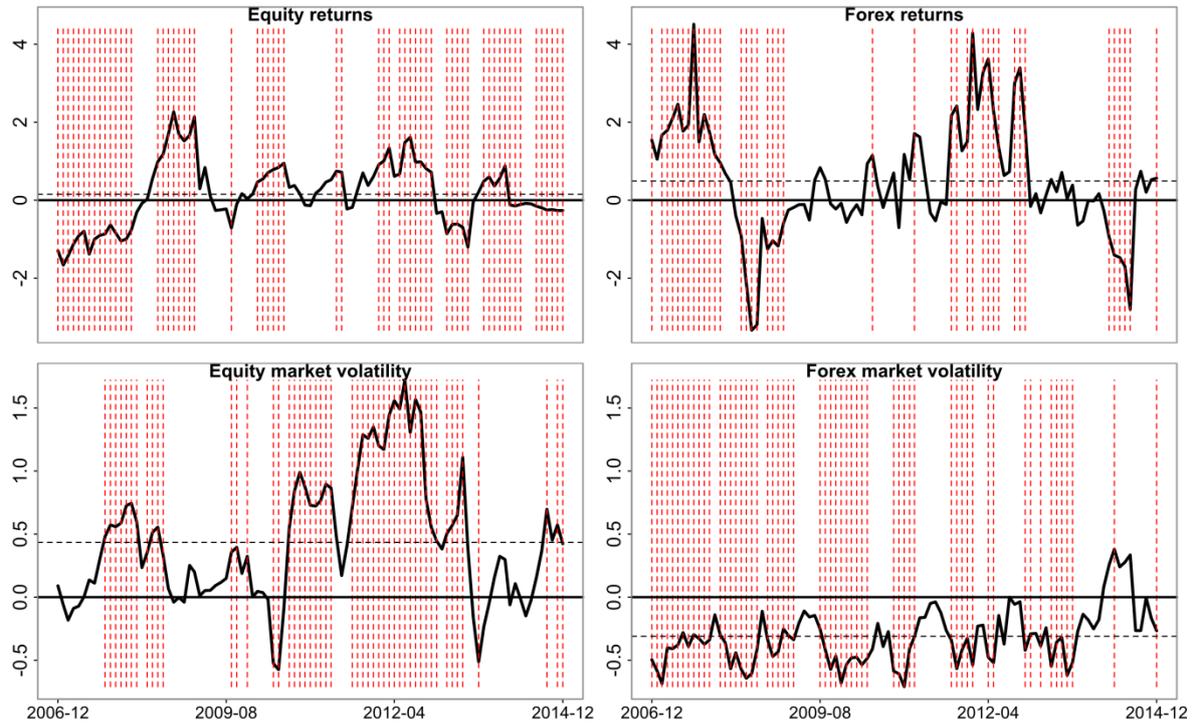

Figure 5: Return and volatility coefficients
*Note: Vertical lines are drawn when the estimated coefficients were statistically significant at the 5% level. The dashed horizontal line is the average value of the spatial probit coefficient.*

Within our approach we control for conditional heteroscedasticity as the standardised residuals used when testing for Granger causality pass the test of no linear autocorrelation and conditional heteroscedasticity. Therefore positive and significant coefficients loaded on the equity market volatility give support to the contagion hypothesis, i.e. during times of higher market uncertainty, the probability of a return spillover increases.

The average size of the foreign exchange volatility is similar to that of the equity market volatility, thus with smaller absolute coefficients foreign market volatility appears to have smaller effect on return spillovers. On the other hand, our results consistently across almost all sub-sample periods suggest that foreign exchange volatility is detrimental for the establishment of return spillover. This is in-line with an explanation that increased volatility on the foreign exchange market is perceived as an increased risk factor for investors, thus leading to home bias effects as investors tend to prefer domestic rather than foreign assets (e.g. Caporale et al., 2015).

### 4.3.2   Is market development important for return spillovers?

Our model also included two variables for controlling the size and hence possible strength of the market in the transmission of information. As can be seen in Figure 6, the sheer size of the market is an important determinant suggesting that the larger the markets is, the more likely it is that we will observe return spillovers from that market. The effect of the relative size of the market to that of the GDP was not stable across the observed period. During the financial crisis, relatively larger markets were less influential (note that we are already controlling for the market capitalization). After the crisis period, more developed markets were more influential within our network of return spillovers.

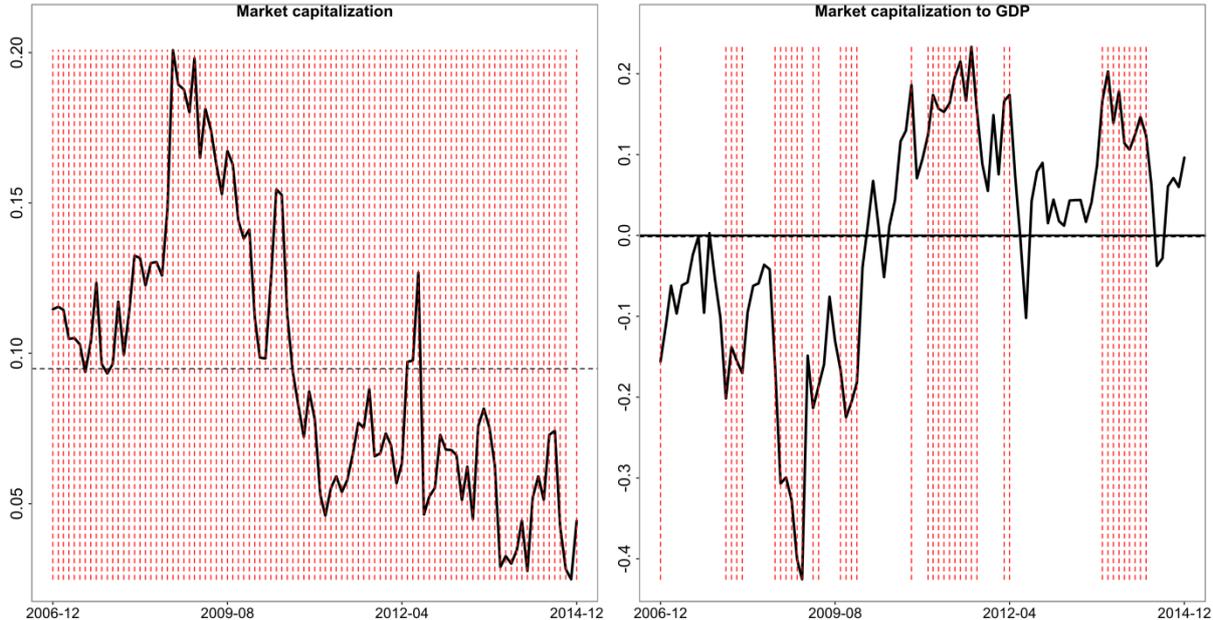

Figure 6: Market capitalization and market capitalization to GDP coefficients
*Note: Vertical lines are drawn when the estimated coefficients were statistically significant at the 5% level. The dashed horizontal line is the average value of the spatial probit coefficient.*

The transmission of information from developed to frontier markets is observed in most samples prior to 2014 (Figure 7). Interestingly, this effect has reversed at the end of the sample, where the linkage from the developed to frontier markets declined. The direction from developed to emerging markets was even less frequently significant and hovered around zero.

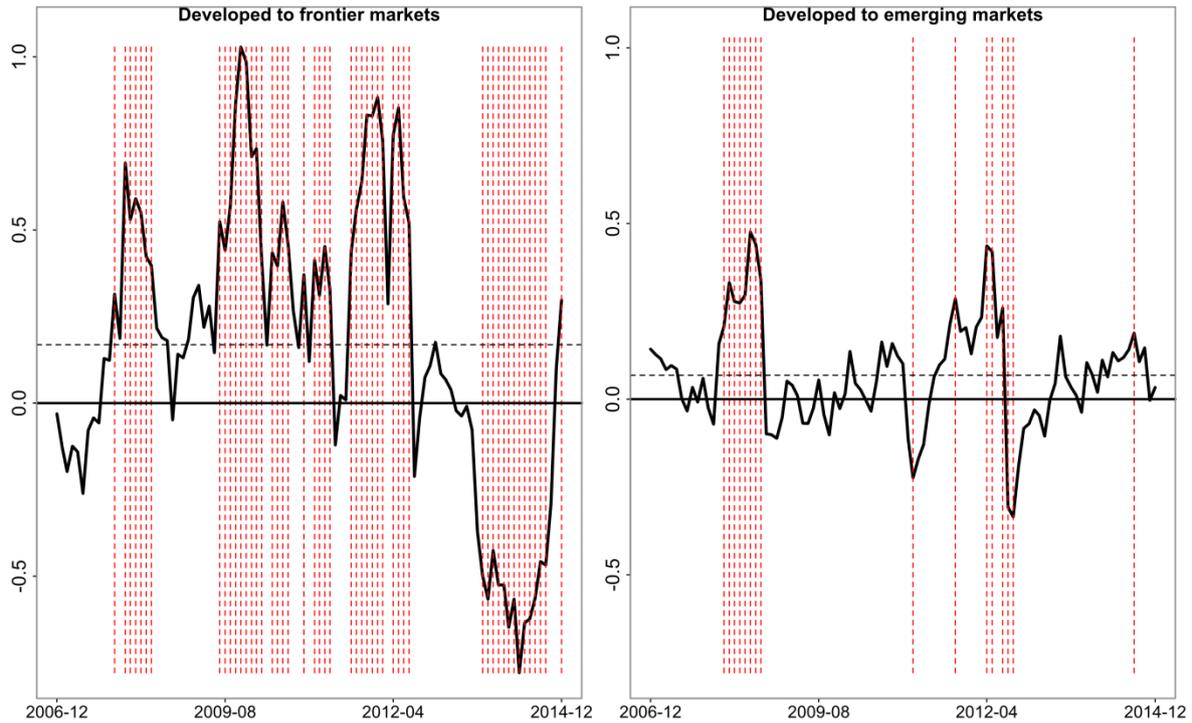

Figure 7: Developed to frontier and developed to emerging markets coefficients
*Note: Vertical lines are drawn when the estimated coefficients were statistically significant at the 5% level. The dashed horizontal line is the average value of the spatial probit coefficient.*

### 4.3.3 Do spatial factors influence return spillovers?

Looking at the results in Table 2, it is obvious, that the significance of both the spatial autocorrelation and temporal proximity is strongly supported. The time evolution of the respective coefficients is depicted in Figure 8 and Figure 9.

The coefficient for temporal distance is strictly negative and significant in all rolling windows. The further the markets trade, the less likely it is that they are connected with an edge; i.e. that returns spillovers happen. The temporal distance to the US market is significant in 84 out of 97 cases (always being negative). Thus, the US can be seen to have an important role in world stock markets, even though the mutual distance remains dominant. The spatial autocorrelation coefficient is also almost always significant (96 out of 97 cases), and is always positive – this can in turn be interpreted as strong evidence for preferential attachment, where the more connections a vertex has, the more likely it is to form new ones.

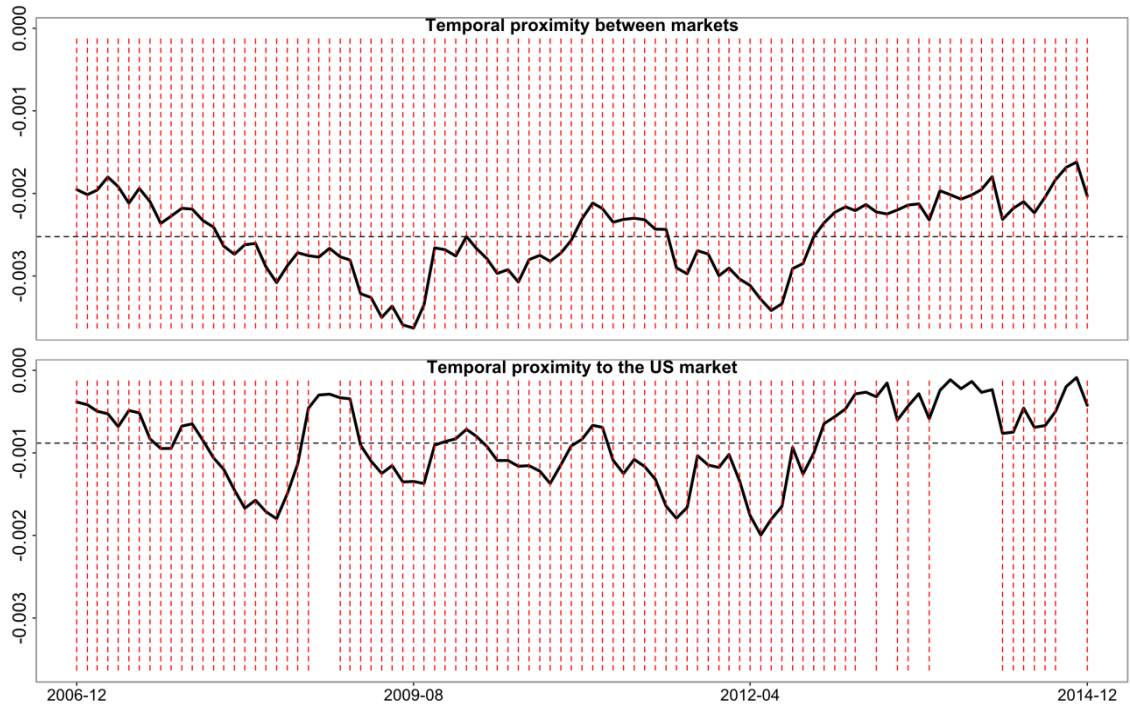

Figure 8: Temporal proximity coefficients: temporal distance to the US market and temporal distance between markets coefficients
*Note: Vertical lines are drawn when the estimated coefficients were statistically significant at the 5% level. The dashed horizontal line is the average value of the spatial probit coefficient.*

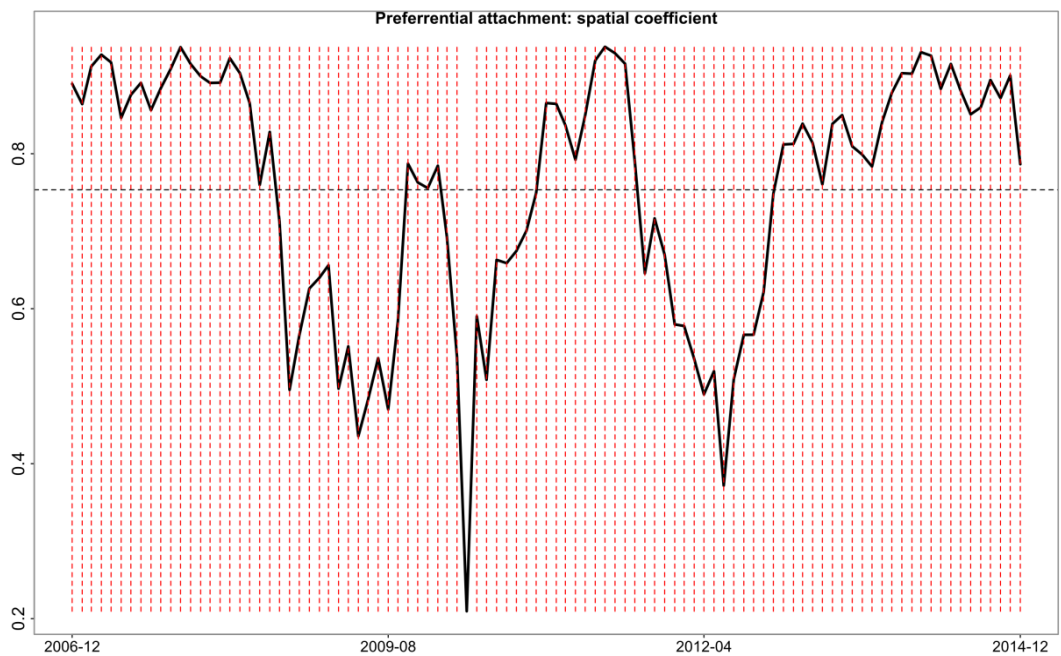

Figure 9: Preferential attachment coefficient: the spatial coefficient
*Note: Vertical lines are drawn when the estimated coefficients were statistically significant at the 5% level. The dashed horizontal line is the average value of the spatial probit coefficient.*

# 5 Conclusion

By constructing a rolling window analysis of Granger causality networks, we have explored the ensuing structures and fitted spatial probit model to explain the way the edges, representing empirical relationships, are constructed. We have confirmed some of the results of Výrost et al. (2015) using a much larger set of equity markets, while we also provided new evidence on determinants of equity market return spillovers.

Our observation of market centralities revealed that the density of return spillovers decreased, i.e. after the crisis the markets are less inter-connected. At the same time, the return spillover between equity markets are quite robust, as on average over 70% of spillovers (edges in the Granger causality network) are retained even after a period of 12 months. Although we do not directly address the effect of such disengagement of equity market relationships, we could also interpret such stability of spillovers as the good news for international equity portfolio diversification, as it makes the portfolio allocation more predictable.

We further observe a peculiar position of the Chinese stock market which not only influenced a small number of other markets around the world, but was also influenced by a small number of markets. This shows, how (relative to other market in our sample) segmented the behaviour of the Chinese market was in the last years.

Interestingly, one would expect positive effect of equity returns, i.e. the higher the return on one market the more likely is a return spillover to other market. Although we found evidence that such positive effects are likely, the reverse, negative effect is also possible and could be explained by flight of investors from one market (e.g. declining) to another (increasing) market. Similar effects were also observed with return on the foreign exchange markets.

Numerous studies before confirmed, that an increased volatility on the market also increases the co-movement between markets. We confirm this within a Granger causality network framework in that our spatial probit models revealed that during times of higher market uncertainty, the probability of a return spillover increases. However, increased volatility on the forex market has the opposite effect, which is in line with the home bias hypothesis, where risk-averse investors restrain from foreign investments when foreign exchange volatility is high.

We further provide strong evidence for preferential attachment effects – that is, the probability of a given market to propagate return spillovers to a new market depends endogenously and positively on the existing number of return spillovers from that market. This result has rather methodological implications as it shows, how important it is to control for the inherent endogeneity between return spillovers.

Our analysis of return spillovers also revealed that existence of return spillovers is related to the size of the market under scrutiny. The larger the market capitalization, the more likely are return spillovers originating from that market to others around the world.

Our results also show strong support for the significance of the temporal proximity effect, i.e. the temporal distance between closing trading hours matters. The closer the closing hours, the more likely is the return spillover from a market which closes first. Although this result is not unexpected, our study is the first which shows such effect on a larger scale of 40 developed, emerging, and frontier markets around the world. When using daily data, such findings hint on the care that should be taken when considering conducting an analysis in an international setting on markets with non-overlapping trading windows. As the temporal effects are highly significant, using appropriate information sets at each individual market is paramount.

The temporal proximity effect is important not only between a given pair of markets, but also with respect to the US market alone. While the importance of the closeness to the US market was to be expected, the evidence for the relation of Granger causality and the closeness of individual market closing times is notable in portfolio management setting, as the return spillovers literally do "travel around the world", but dies out as the temporal distance between markets increases.

How can this be used in portfolio investment analysis? It seems, that a choice of two markets, which have similar trading hours bears additional costs, as a return spillover from one market to another is more likely. Note, that temporal proximity is not equivalent to geographical proximity, therefore a behavioural explanation is most natural for this phenomenon: the fundamental problem might be that there might not be simply enough time for investors to rationally evaluate, whether this shock on one market will in reality effect the underlying economy on the other market.

**Acknowledgement**

This work was supported by the Slovak Research and Development Agency under the contract No. APVV-0666-11 and No. APVV-14-0357. Authors also appreciate the support provided from the Slovak Grant Agency for Science (VEGA project No. 1/0392/15).

## References

[1] Aloui, R., Aïssa, M.S.B., Nguyen, D.K. 2011. Global financial crisis, extreme interdependences, and contagion effects: the role of economic structure? Journal of Banking and Finance, vol. 35, no. 1, p. 130–41.

[2] Arshanapalli, B., Doukas, J. 1993. International Stock Market Linkages: Evidence from the Pre- and Post-October 1987 period. Journal of Banking & Finance, vol. 17, no. 1, p. 193–208.

[3] Barabási, A.L., Albert, R. 1999. Emergence of Scaling in Random Networks. Science, vol. 286, no. 5439, p. 509–12.

[4] Barabási, A.L., Gulbahce, N., Loscalzo, J. 2011. Network medicine: a network-based approach to human disease. Nature Reviews Genetics, vol. 12, p. 56–68.

[5] Billio, M., Getmansky, M., Lo, A.W., Pelizzon, L. 2012. Econometric measures of connectedness and systemic risk in the finance and insurance sectors. Journal of Financial Economics, vol. 104, no. 3, p. 535–59.

[6] Bohn, H., Tesar, L. L. 1996. U.S. Equity Investment in Foreign Markets: Portfolio Rebalancing or Return Chasing? The American Economic Review, vol. 86, no. 2, p. 77–81.

[7] Boldi, P., Vigna, S. 2014. Axioms for Centrality. Internet Mathematics, vol. 10, no. 3-4, p. 222–62.

[8] Bollerslev, T. 1986. Generalized Autoregressive Conditional Heteroskedasticity. Journal of Econometrics, vol. 31, no. 3, p. 307–27.

[9] Bonanno, G., Vandewalle, N., Mantegna, R.N. 2000. Taxonomy of Stock Market Indices. Physical Review E, vol. 62, R7615.

[10] Bullmore, E., Sporns, O. 2009. Complex brain networks: graph theoretical analysis of structural and functional systems. Nature Reviews Neuroscience, vol. 10, p. 186–98.

[11] Caporale, G.M., Ali, F.M., Spagnolo, N. 2015. Exchange rate uncertainty and international portfolio flows: A multivariate GARCH-in-mean approach. Journal of International Money and Finance, vol. 54, June, p. 70–92.

[12] Cappiello, L., Engle, R.H., Sheppard, K. 2006. Asymmetric dynamics in the correlations of global equity and bond returns. Journal of Financial Econometrics, vol. 4, no. 4, p. 537–72.

[13] Cenedese, G., Payne, R., Sarno, L., Valente, G. 2015. What do stock markets tell us about exchange rates? Bank of England, Staff working paper No. 537.

[14] Cheung, Y.-W., Ng, L.K. 1996. A causality-in-variance test and its application to financial market prices. Journal of Econometrics, vol. 72, no. 1-2, p. 33–48.

[15] Choi, P., Nam, K. 2008. Asymmetric and leptokurtic distribution for heteroscedastic asset returns: The SU-normal distribution. Journal of Empirical Finance, vol. 15, no. 1, p. 41–63.

[16] Coelho, R., Gilmore, C.G., Lucey, B., Richmond, P., Hutzler, S. 2007. The evolution of interdependence in world equity markets—Evidence from minimum spanning trees. Physica A: Statistical Mechanics and its Applications, vol. 376, p. 455–66.

[17] Curme, Ch., Tumminello, M., Mantegna, R.N., Stanley, H.E., Kennet, D.Y. 2014. Emergence of statistically validated financial intraday lead-lag relationships. arXiv:1401.0462v1 [q-fin.ST].

[18] Diebold, F.X., Yilmaz, K. 2014. On the network topology of variance decompositions: Measuring the connectedness of financial firms. Journal of Econometrics, vol. 182, no. 1, p. 119–34.

[19] Ding, Z., Engle, R.F., Granger, C.W.J. 1993. A Long Memory Property of Stock Market Returns and a New Model. Journal of Empirical Finance, vol. 1, no. 1, p. 83–106.

[20] Dorogovtsev, S.N., Mendes, J.F.F. 2003. Evolution of Networks: From Biological Nets to the Internet and WWW. Oxford University Press, Oxford, 2003.

[21] Engle, R.F., Ng, V.K. 1993. Measuring and Testing the Impact of News On Volatility. Journal of Finance, vol. 48, no. 5, p. 1749–78.

[22] Eryiğit, M., Eryiğit, R. 2009. Network structure of cross/correlations among the world market indices. Physica A: Statistical Mechanics and its Applications, vol. 388, no. 17, p. 3551–62.

[23] Freeman, L.C. 1979. Centrality in social networks conceptual clarification. Social Networks, vol. 1, no. 3, p. 1978–9.

[24] Ghalanos, A. 2012a. rmgarch: Multivariate GARCH models. R package version 1.2-9.

[25] Ghalanos, A. 2012b. rugarch: Univariate GARCH models. R package version 1.3-6.

[26] Gilmore, C.G., Lucey, B.M., Boscia, M. 2008. An ever-closer union? Examining the evolution of linkages of European equity markets via minimum spanning trees. Physica A: Statistical Mechanics and its Applications, vol. 387, no. 25, p. 6319–29.

[27] Girvan, M., Newman, M.E.J. 2002. Community structure in social and biological networks. Proceedings of the National Academy of Sciences of the United States of America, vol. 99, no. 2, p. 7821–26.

[28] Glosten, L.R., Jagannathan, R., Runkle, D.E. 1993. On the Relationship Between the Expected Value and the Volatility of the Nominal Excess Return on Stocks. Journal of Finance, vol. 48, no. 5, p. 1779–1801.

[29] Granger, C.W.J. 1969. Investigating Causal Relations by Econometric Models and Cross-spectral Methods. Econometrica, vol. 37, no. 3, p. 424–38.

[30] Granger, C.W.J. 1980. Testing for causality: A personal viewpoint. Journal of Economic Dynamics and Control, vol. 2, 329–52.

[31] Grubel, H. 1968. Internationally diversified portfolios: Welfare gains and capital flows. American Economic Review, vol. 58, no. 5, p. 1299–14.

[32] Hardouvelis, G., Malliaropoulos, D., Priestley, R. 2006. EMU and European Equity Market Integration. Journal of Business, vol. 79, no. 1, p. 365–92.

[33] Hau, H., Rey, H. 2006. Exchange rates, equity prices, and capital flows. Review of Financial Studies, vol. 19, no. 1, p. 273–17.

[34] Hentschel, L. 1995. All in the family: Nesting symmetric and asymmetric GARCH models. Journal of Financial Economics, vol. 39, no. 1, p. 71–104.

[35] Higgins, M.L., Bera, A.K. 1992. A Class of Nonlinear ARCH Models. International Economic Review, vol. 33, no. 1, p. 137–58.

[36] Hong, Y. 2001. A test for volatility spillover with application to exchange rates. Journal of Econometrics, vol. 103, no. 1-2, p. 183–224.

[37] Jaffe, J., Westerfield, R. 1985. The Week-End Effect in Common Stock Returns: The International Evidence. Journal of Finance, vol. 40, no. 2, 433–54.

[38] Johnson, N.L. 1949a. Systems of Frequency Curves Generated by Method of Translation. Biometrika, vol. 36, no. 1/2, p. 149–76.

[39] Johnson, N.L. 1949b. Bivariate Distributions Based on Simple Translation Systems. Biometrika, vol. 36, no. 3/4, p. 297–304.

[40] Kullmann, L., Kertész, J., Kaski, K. 2002. Time-dependent cross-correlations between different stock returns: A directed network of influence. Physical Review E, vol. 66, 026125.

[41] Lee, G.J., Engle, R.F. 1999. A permanent and transitory component model of stock return volatility. In Cointegration Causality and Forecasting: A Festschrift in Honor of Clive W. J. Granger, pages 475–97. Oxford University Press.

[42] LeSage, J. P., Pace, R. K. 2010. Introduction to spatial econometrics. CRC press, Boca Raton, 2010.

[43] LeSage, J.P. 2000. Bayesian estimation of limited dependent variable spatial autoregressive models. Geographical Analysis, vol. 32, no. 1, p. 19–35.

[44] Lin, J.W., McLeod, A.I. 2006. Improved Peña-Rodríguez Portmanteau test. Computational Statistics & Data Analysis, vol. 51, no. 3, p. 1731–8.

[45] Longin, F., Solnik, B. 1995. Is the correlation in international equity returns constant: 1960–1990? Journal of International Money and Finance, vol. 14, no. 1, p. 3–26.

[46] Lu, F.B., Hong, Y.-M., Wang, S.-Y., Lai, K.-K., Liu, J. 2014. Time-varying Granger causality tests for applications in global crude oil markets. Energy Economics, vol. 42, p. 289–98.

[47] Lyócsa, Š., Výrost, T., Baumöhl, E. 2012. Stock market networks: The dynamic conditional correlation approach. Physica A: Statistical Mechanics and its Applications, vol. 391, no. 16, p. 4147–58.

[48] Mantegna, R.N. 1999. Hierarchical structure in financial markets. The European Physical Journal B, vol. 11, no. 1, p. 193–7.

[49] Mantegna, R.N., Stanley, H.E. 1999. Introduction to Econophysics: Correlations and Complexity in Finance, Cambridge University Press, Cambridge, 1999.

[50] McMillen, D.P. 1992. Probit with spatial autocorrelation. Journal of Regional Science, vol. 32, no. 3, p. 335–48.

[51] Mink, M. 2015. Measuring stock market contagion: Local or common currency returns? Emerging Markets Review, vol. 22, March, p. 18–24.

[52] Mylonidis, N., Kollias, Ch. 2010. Dynamic European stock market convergence: Evidence from rolling cointegration analysis in the first euro-decade. Journal of Banking and Finance, vol. 34, no. 9, p. 2056–64.

[53] Nelson, D.B. 1991. Conditional Heteroskedasticity in Asset Returns: A New Approach. Econometrica, vol. 59, no. 2, p. 347–70.

[54] Newey, W.K., West, K.D. 1994. Automatic Lag Selection in Covariance Matrix Estimation. The Review of Economic Studies, vol. 61, no. 4, p. 631–53.

[55] Onnela, J.P., Chakraborti, A., Kaski, K., Kertész, J., 2003a. Dynamic Asset trees and Black Monday. Physica A: Statistical Mechanics and its Applications, vol. 324, no. 1, p. 247–52.

[56] Onnela, J.P., Chakraborti, A., Kaski, K., Kertész, J., Kanto, A. 2003b. Dynamics of market correlations: Taxonomy and portfolio analysis. Physical Review E, vol. 68, 056110.

[57] Peña, D., Rodríguez, J., 2006. The log of the determinant of the autocorrelation matrix for testing goodness of fit in time series. Journal of Statistical Planning and Inference, vol. 136, no. 8, p. 2706–18.

[58] Schwartz, G. 1978. Estimating the Dimension of a Model. Annals of Statistics, vol. 6, no. 2, p. 461–4.

[59] Solnik, B. 1974. Why not diversify internationally rather than domestically? Financial Analysts Journal, vol. 30, no. 4, p. 48–54.

[60] Song, D.-M., Tumminello, M., Zhou, W.-X., Mantegna, R.N. 2011. Evolution of worldwide stock markets, correlation structure and correlation based graphs. Physical Review E, vol. 84, 026108.

[61] Tabak, B.M., Serra, T.R., Cajueiro, D.O. 2010. Topological properties of stock market networks: The case of Brazil. Physica A: Statistical Mechanics and its Applications, vol. 389, no. 16, p. 3240–9.

[62] Taylor, S.J. 1986. Modeling Financial Time Series. John Wiley & Sons Ltd, Chichester, 1986.

[63] Trancoso, T. 2014. Emerging markets in the global economic network: Real(ly) decoupling? Physica A: Statistical Mechanics and its Applications, vol. 395, p. 499–510.

[64] Tumminello, M., Di Mateo, T., Aste, T., Mantegna, R.N. 2007. Correlation based networks of equity returns sampled at different time horizons. The European Physical Journal B, vol. 55, no. 2, p. 209–17.

[65] Výrost, T., Lyócsa, Š., Baumöhl, E. 2015. Granger causality stock market networks: Temporal proximity and preferential attachment. Physica A: Statistical Mechanics and its Applications, vol. 427, p. 262–76.

[66] Zakoian, J-M. 1994. Threshold Heteroskedastic Models. Journal of Economic Dynamics and Control, vol. 18, no. 5, p. 931–55.

Appendix A

| Code | Country | Index |
|---|---|---|
| *Frontier markets* | | |
| **AR** | Argentina | MSCI ARGENTINA |
| **HR** | Croatia | CROATIA CROBEX |
| **EE** | Estonia | OMX TALLINN (OMXT) |
| **RO** | Romania | ROMANIA BET |
| **SL** | Slovenia | SLOVENIA-DS Market |
| *Emerging markets* | | |
| **BR** | Brazil | MSCI BRAZIL |
| **CZ** | Czech Republic | PRAGUE SE PX |
| **HU** | Hungary | BUDAPEST |
| **CN** | China | SHANGHAI SE COMPOSITE |
| **IN** | India | S&P BSE NATIONAL 200 |
| **ID** | Indonesia | IDX COMPOSITE |
| **MY** | Malaysia | DJGL MALAYSIA |
| **MX** | Mexico | MEXICO IPC (BOLSA) |
| **PL** | Poland | WARSAW GENERAL INDEX 20 |
| **RU** | Russia | RUSSIA-DS Market |
| **ZA** | South Africa | SOUTH AFRI-DS Market |
| **KR** | Republic of Korea | KOREA SE KOSPI 200 |
| **TH** | Thailand | BANGKOK S.E.T. |
| **TR** | Turkey | TURKEY-DS Market |
| *Developed markets* | | |
| **AU** | Australia | ASX 200 |
| **AT** | Austria | ATX - AUSTRIAN TRADED INDEX |
| **BE** | Belgium | BEL ALL SHARE |
| **CA** | Canada | S&P/TSX Composite index |
| **DK** | Denmark | DENMARK-DS Market |
| **FI** | Finland | OMX HELSINKI 25 |
| **FR** | France | FRANCE CAC 40 |
| **DE** | Germany | DAX 30 PERFORMANCE |
| **GR** | Greece | GREECE-DS Market |
| **IE** | Ireland | IRELAND SE OVERALL |
| **IT** | Italy | MSCI ITALY |
| **JP** | Japan | NIKKEI 225 STOCK AVERAGE |
| **NL** | Netherlands | AMSTERDAM MIDKAP |
| **HK** | Hong Kong | HANG SENG |
| **NO** | New Zealand | AEX ALL SHARE |
| **PT** | Portugal | PORTUGAL PSI-20 |
| **ES** | Spain | IBEX 35 |
| **SE** | Sweden | OMX STOCKHOLM 30 |
| **CH** | Switzerland | SSMI |
| **UK** | United Kingdom | FTSE ALL SHARE |
| **US** | United States of America | RUSSELL 2000 |

Appendix B

|  | Equity returns | | Equity volatility | | FX return | | FX volatility | | MC | | MC/GDP | |
|---|---|---|---|---|---|---|---|---|---|---|---|---|
|  | Mean | SD | Mean | SD | Mean | SD | Mean | SD | Mean | SD | Mean | SD |
| *Frontier markets* | | | | | | | | | | | | |
| AR | 0.318 | 0.540 | 0.943 | 0.260 | 0.139 | 0.156 | 0.695 | 0.756 | 24.637 | 0.314 | 2.523 | 0.530 |
| HR | 0.021 | 0.360 | 0.889 | 0.532 | 0.003 | 0.099 | 0.981 | 0.304 | 24.035 | 0.326 | 3.775 | 0.362 |
| EE | 0.100 | 0.404 | 0.968 | 0.368 | -0.002 | 0.095 | 0.977 | 0.311 | 21.711 | 0.389 | 2.449 | 0.399 |
| RO | 0.065 | 0.403 | 0.935 | 0.424 | 0.030 | 0.137 | 0.976 | 0.328 | 23.911 | 0.336 | 2.584 | 0.395 |
| SL | 1.378 | 3.973 | 0.382 | 0.942 | -0.002 | 0.095 | 0.977 | 0.311 | 23.040 | 0.482 | 2.992 | 0.485 |
| *Emerging markets* | | | | | | | | | | | | |
| BR | 0.059 | 0.262 | 0.925 | 0.400 | 0.022 | 0.162 | 0.931 | 0.372 | 27.763 | 0.212 | 4.081 | 0.289 |
| CZ | -0.016 | 0.247 | 0.932 | 0.437 | -0.001 | 0.131 | 0.976 | 0.340 | 24.544 | 0.207 | 3.109 | 0.260 |
| HU | 0.015 | 0.313 | 0.956 | 0.357 | 0.024 | 0.149 | 0.981 | 0.322 | 23.971 | 0.283 | 2.922 | 0.318 |
| CN | 0.173 | 0.625 | 0.961 | 0.347 | 0.003 | 0.099 | 0.963 | 0.369 | 29.019 | 0.178 | 4.197 | 0.472 |
| IN | 0.148 | 0.321 | 0.938 | 0.390 | 0.044 | 0.113 | 1.001 | 0.261 | 27.817 | 0.203 | 4.328 | 0.314 |
| ID | 0.217 | 0.327 | 0.979 | 0.292 | 0.038 | 0.113 | 0.897 | 0.418 | 26.290 | 0.459 | 3.693 | 0.301 |
| MY | 0.110 | 0.215 | 0.955 | 0.378 | -0.011 | 0.065 | 1.005 | 0.194 | 26.569 | 0.296 | 4.943 | 0.226 |
| MX | 0.118 | 0.238 | 0.941 | 0.371 | 0.032 | 0.118 | 0.959 | 0.388 | 26.728 | 0.227 | 3.632 | 0.235 |
| PL | 0.005 | 0.242 | 0.969 | 0.309 | 0.018 | 0.181 | 0.964 | 0.373 | 25.783 | 0.185 | 3.494 | 0.298 |
| RU | 0.035 | 0.322 | 0.890 | 0.480 | 0.056 | 0.165 | 0.694 | 0.319 | 27.509 | 0.245 | 3.934 | 0.444 |
| ZA | 0.108 | 0.164 | 0.945 | 0.315 | 0.071 | 0.151 | 0.983 | 0.283 | 27.169 | 0.119 | 5.208 | 0.250 |
| KR | 0.068 | 0.208 | 0.957 | 0.379 | 0.024 | 0.153 | 0.035 | 0.305 | 27.598 | 0.220 | 4.545 | 0.224 |
| TH | 0.130 | 0.284 | 0.993 | 0.272 | -0.017 | 0.064 | 0.898 | 0.494 | 26.161 | 0.437 | 4.340 | 0.347 |
| TR | 0.134 | 0.349 | 0.980 | 0.238 | 0.063 | 0.138 | 0.942 | 0.333 | 26.212 | 0.246 | 3.520 | 0.318 |
| *Developed markets* | | | | | | | | | | | | |
| AU | 0.026 | 0.185 | 0.965 | 0.357 | 0.033 | 0.152 | 0.959 | 0.392 | 27.808 | 0.164 | 4.590 | 0.285 |
| AT | -0.016 | 0.274 | 0.954 | 0.385 | -0.002 | 0.095 | 0.977 | 0.311 | 25.319 | 0.417 | 3.179 | 0.432 |
| BE | 0.024 | 0.229 | 0.962 | 0.345 | -0.002 | 0.095 | 0.977 | 0.311 | 26.352 | 0.195 | 4.028 | 0.263 |
| CA | 0.042 | 0.171 | 0.904 | 0.495 | 0.007 | 0.097 | 0.963 | 0.343 | 28.234 | 0.178 | 4.736 | 0.230 |
| DK | 0.101 | 0.258 | 0.954 | 0.372 | -0.003 | 0.095 | 0.977 | 0.312 | 26.062 | 0.159 | 4.168 | 0.243 |
| FI | 0.053 | 0.268 | 0.976 | 0.327 | -0.002 | 0.095 | 0.977 | 0.311 | 25.820 | 0.370 | 4.142 | 0.377 |
| FR | 0.001 | 0.198 | 0.970 | 0.344 | -0.002 | 0.095 | 0.977 | 0.311 | 28.281 | 0.154 | 4.261 | 0.204 |
| DE | 0.086 | 0.214 | 0.961 | 0.356 | -0.002 | 0.095 | 0.977 | 0.311 | 27.997 | 0.150 | 3.729 | 0.209 |
| GR | -0.110 | 0.322 | 0.966 | 0.235 | -0.002 | 0.095 | 0.977 | 0.311 | 25.005 | 0.647 | 3.124 | 0.577 |
| IE | -0.002 | 0.291 | 0.945 | 0.419 | -0.002 | 0.095 | 0.977 | 0.311 | 25.249 | 0.341 | 3.132 | 0.362 |
| IT | -0.047 | 0.223 | 0.971 | 0.323 | -0.002 | 0.095 | 0.977 | 0.311 | 26.952 | 0.384 | 3.672 | 0.387 |
| JP | 0.028 | 0.257 | 0.964 | 0.337 | -0.005 | 0.116 | 0.987 | 0.250 | 28.958 | 0.098 | 4.231 | 0.172 |
| NL | 0.016 | 0.223 | 0.936 | 0.434 | -0.002 | 0.095 | 0.977 | 0.311 | 27.176 | 0.190 | 4.360 | 0.270 |
| HK | 0.075 | 0.259 | 0.943 | 0.441 | 0.000 | 0.003 | 0.952 | 0.424 | 27.697 | 0.097 | 6.119 | 0.157 |
| NO | 0.042 | 0.249 | 0.923 | 0.422 | 0.007 | 0.127 | 0.979 | 0.317 | 26.204 | 0.199 | 3.953 | 0.314 |
| PT | -0.022 | 0.244 | 0.974 | 0.283 | -0.002 | 0.095 | 0.977 | 0.311 | 25.108 | 0.230 | 3.531 | 0.245 |
| ES | 0.011 | 0.228 | 0.987 | 0.300 | -0.002 | 0.095 | 0.977 | 0.311 | 27.776 | 0.172 | 4.386 | 0.204 |
| SE | 0.064 | 0.221 | 0.959 | 0.367 | 0.000 | 0.131 | 0.975 | 0.337 | 26.929 | 0.208 | 4.607 | 0.275 |
| CH | 0.026 | 0.184 | 0.955 | 0.390 | -0.034 | 0.090 | 0.981 | 0.300 | 27.708 | 0.093 | 5.240 | 0.210 |
| UK | 0.034 | 0.165 | 0.954 | 0.397 | 0.019 | 0.118 | 0.959 | 0.364 | 28.718 | 0.156 | 4.761 | 0.206 |
| US | 0.083 | 0.221 | 0.950 | 0.410 | -0.002 | 0.095 | 0.977 | 0.311 | 30.457 | 0.133 | 4.695 | 0.149 |

*Note: The Table reports basic statistics from data used in the spatial probit model. Returns are calculated across the whole rolling window. The standard deviation (SD) of returns is the realized volatility from daily squared returns across the whole rolling window. For both, Market capitalization (MC) and Market capitalization to GDP (MC/GDP) we took the average from their logarithms.*